\documentclass[11pt,twoside]{article}
\usepackage{./asp2014}

\aspSuppressVolSlug
\resetcounters

\begin{document}

\title{Mass-radius relation of strongly magnetized white dwarfs}
\author{Prasanta Bera$^*$ and Dipankar Bhattacharya,
\affil{Inter University Centre for Astronomy and Astrophysics, Post Bag 4, Pune 411007, India; $^*$\email{pbera@iucaa.in}}}

\begin{abstract}
We study the strongly magnetized white dwarf configurations in a self-consistent manner as a progenitor of the over-luminous type-Ia supernovae. We compute static equilibria of white dwarf stars containing a strong magnetic field and present the modification of white dwarf mass-radius relation caused by the magnetic field. From a static equilibrium study, we find that a maximum white dwarf mass of about 1.9 M$_\odot$ may be supported if the interior poloidal field is as strong as approximately $10^{10}$ T. On the other hand if the field is purely toroidal the maximum mass can be more than 5 M$_\odot$. All these modifications are mainly from the presence of Lorenz force. The effects of i) modification of equation of state due to Landau quantization, ii) electrostatic interaction due to ions, iii) general relativistic calculation on the stellar structure and, iv) field geometry are also considered. These strongly magnetised configurations are sensitive to magnetic instabilities where the perturbations grow at the corresponding Alfven time scales.
\end{abstract}

\section{Introduction}
White dwarfs are electron degenerate compact systems and this degeneracy pressure force is able to sustain the inward gravitational force only when the white dwarf mass is below the Chandrasekhar limit \citep{Chandrasekhar1931}. In general, the limiting mass value is 1.44~M$_\odot$ but this value can extend in presence of rotation/magnetic field \citep{Ostriker+Hartwick1968}. White dwarfs are considered to be the progenitor of the type-Ia supernovae (SNIa). In a simplified scenario, a white dwarf in a binary system can increase its mass by accretion from the companion and may exceed the limiting mass to collapse forming the SNIa. The characteristic light curve, the indicator of the limiting mass, has been used to measure the distances of galaxies. A few Recently observed SNIa are over-luminous and suggest the presence of white dwarfs with mass more than 2~M$_\odot$ \citep{Howell+2006,Hicken+2007}. The presence of strong internal magnetic field is one of the possible proposals (the others being rapid rotation, the merger of two white dwarfs etc.) to support a massive white dwarf \citep{Das+Mukhopadhyay2012}. To study the effects of the strong internal magnetic field on the white dwarf structure, we generate axisymmetric equilibrium structures in a self-consistent method and construct the mass-radius relation of the magnetized white dwarfs. Due to the presence of strong magnetic field,  the electron orbits may be quantized \citep{Lai+Shapiro1991}. We study the effects of the quantized electron orbit on the stellar structure. The effects of general relativistic gravity and electrostatic corrections on the equilibrium structures are also explored. Whether such objects may be found among the White Dwarf pupulation depends on the stability of these equilibrium structures. We perturb equilibrium configurations and study their evolution to examine their stability in the linear and non-linear regime.

\section{Method}
To obtain the self-consistent magnetic white dwarf with Fermi degenerate EoS, we solve the stellar structure equations assuming an axisymmetric structure and the ideal MHD condition in the Newtonian limit,
\begin{align}\label{hydro_eui}
\frac{1}{\rho_0}\mathbf{\nabla} P_0 &=-\mathbf{\nabla} \Phi_g+\frac{1}{\rho_0}\left( \mathbf{j_0}\boldsymbol\times\mathbf{B_0}\right) \\
\mathbf{\nabla}^2\Phi_g &= 4\pi G \rho_0 \\
\nabla \cdot \mathbf{B_0} &= 0,  \\
\nabla \times \mathbf{B_0} &= \mu_0 \mathbf{j_0} .
\label{maxwell}
\end{align}
\\where $P_0$, $\rho_0$, $\Phi_g$, $\mathbf{j_0}$, $\mathbf{B_0}$ and $\mu_0$ are pressure, mass density, gravitational potential, current density, magnetic field and free space permeability respectively. 
These set of equations \ref{hydro_eui}-\ref{maxwell} are transformed to the integral form relating the Fermi energy ($E_F$), gravitational potential and a quantity $M(u)$ dependent on the magnetic flux function ($u$).   The functional form represents the field geometry.
\begin{displaymath}\label{intro_c}
\dfrac{1}{2 m_{\rm H}} E_{\rm F} +\Phi_{\rm g} = M(u)+C, \qquad \textrm{$m_{\rm H}$ : mass of hydrogen atom, $C$ : integration constant}
\end{displaymath}
This integral equation is iteratively used to find the magnetic configuration and the solution is verified with the stellar virial relation which relates the total pressure, gravitational (W) and magnetic energies ($\mathcal{M}$) \citep{Hachisu1986}.

The equilibrium configurations with EoS considering i) energy level quantization due to the magnetic field or ii) electrostatic corrections are obtained in a similar way but only replacing general Fermi degenerate EoS by the Landau quantized EoS \citep{Lai+Shapiro1991} or Salpeter EoS \citep{Salpeter1961}. The configurations with general relativistic gravity are generated using publicly available codes \textsc{lorene}\footnote{\url{http://www.lorene.obspm.fr/}} and \textsc{xns}\footnote{\url{http://www.arcetri.astro.it/science/ahead/XNS/code.html}} with appropriate modifications (e.g. degenerate EoS) suited to the white dwarf structure. 

To study the stability of the equilibrium configurations we evolve the MHD equations in the Newtonian framework assuming the Cowling approximation i.e. the gravitational potential remaining fixed at the equilibrium value. The evolution in the linear regime is obtained by evolving the perturbing variables over the equilibrium configuration. Two-step MacCormack method is used to solve the following set of equations,
\begin{align}\label{hydro_pertb}
\frac{\partial\mathbf{f}}{\partial t} & = \left[-\mathbf{\nabla} \delta P + \frac{\mathbf{\nabla} P_0}{\rho_0}\delta\rho \right] -\frac{1}{\rho_0}\left(\mathbf{J_0}\boldsymbol\times\mathbf{B_0}\right)\delta\rho +  \frac{1}{\rho_0}\left(\mathbf{J_0}\boldsymbol\times\boldsymbol{\beta}\right) \nonumber\\
& + \left[\frac{1}{\rho_0}\left(\nabla\times\boldsymbol{\beta}\right)\times\mathbf{B_0} - \frac{1}{\rho_0^2}\left(\nabla\rho_0\times\boldsymbol{\beta}\right)\times\mathbf{B_0}\right]\\
\frac{\partial\delta\rho}{\partial t} & = -\nabla\cdot\mathbf{f}\\
\frac{\partial \boldsymbol\beta}{\partial t} &= \nabla\times(\mathbf{f}\times\mathbf{B_0})-\frac{\nabla\rho_0}{\rho_0}\times(\mathbf{f}\times\mathbf{B_0})\\ 
\delta P & = \left.\frac{\partial P}{\partial \rho}\right|_0\delta\rho \label{hydro_pertb_end}
\end{align} 
Here, $\delta P, \delta \rho, \delta \mathbf{B}$ are perturbed pressure, density, magnetic field respectively, $\mathbf{f}=\rho_0\mathbf{v}$ ($\mathbf{v}$ : velocity) and $\boldsymbol\beta=\rho_0\delta\mathbf{B}$. Each of these perturbed quantities is further decomposed in azimuthal angle $\phi$ with index $m$, e.g. the perturbed pressure 
$\delta P (t, r, \theta, \phi) = \sum_{m=0}^{m=+\infty}[\delta P^+(t, r, \theta) \cos m\phi + \delta P^-(t, r, \theta) \sin m\phi].$
It is known that the axisymmetric magnetic configurations are unstable against the axial velocity perturbation \citep{Tayler1973, Lander+Jones2011, Braithwaite2006}. To excite the specific unstable modes, we start the evolution of the equilibrium solution with the specific velocity perturbation $\mathbf f \sim \hat{r} \times \nabla Y_{lm}(\theta, \phi)$. On the other hand, the non-linear evolution of the magnetic configurations is computed using the \textsc{pluto}\footnote{\url{http://plutocode.ph.unito.it/}} code for a selected stellar interior region to avoid very low matter density with magnetic field.
\section{Results}
\articlefiguretwo{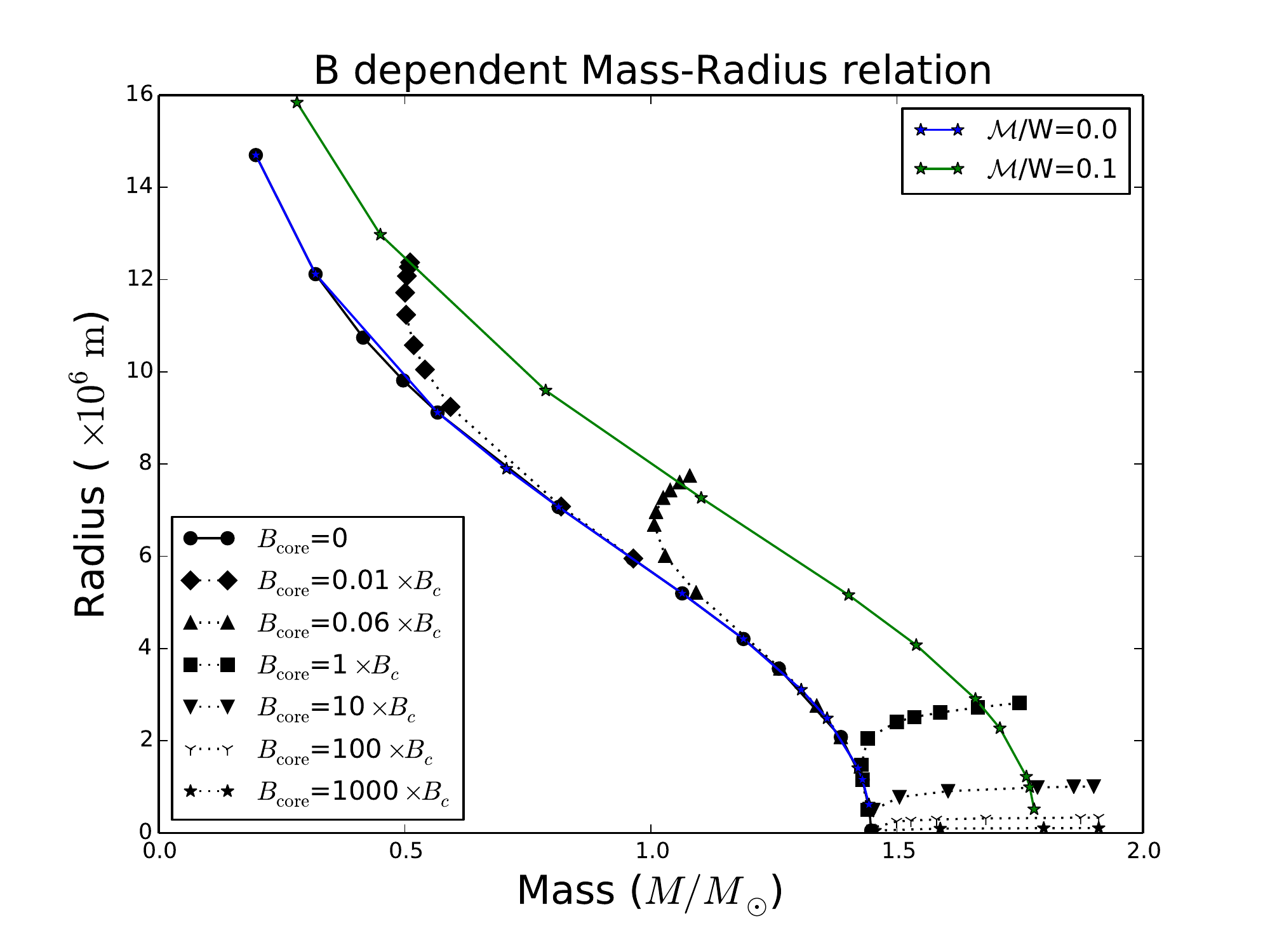}{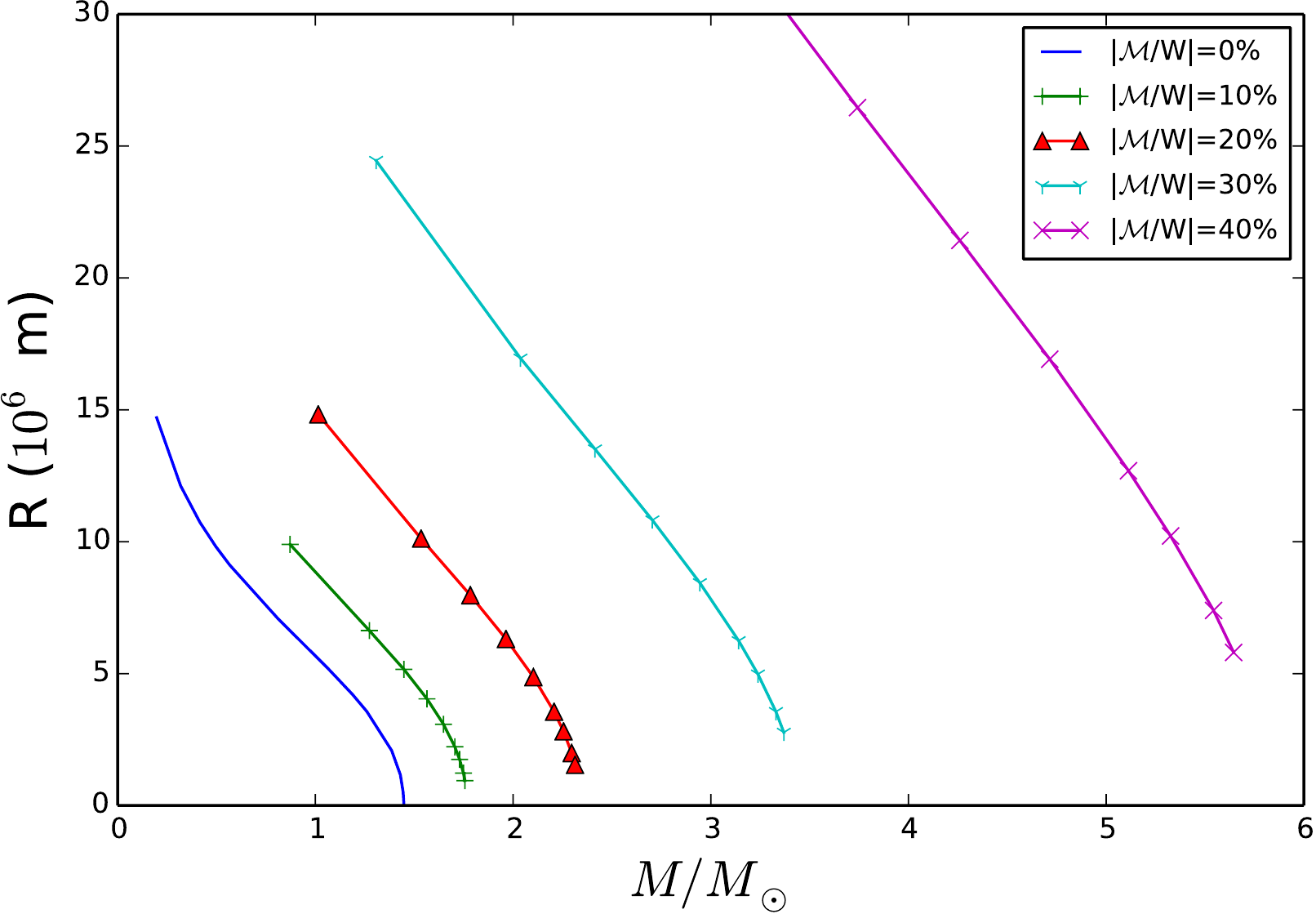}{M_R}{The mass-radius relation of white dwarfs with \emph{Left:} Pure poloidal field.  \emph{Right:} Pure toroidal field. The maximum mass for a pure poloidal field is about 1.9~M$_\odot$ and for pure toroidal field it is more than 5 M$_\odot$.}
The radius of a  non-magnetic white dwarf decreases as the mass increases ($R\propto M^{-1/3}$ in the low mass range) and vanishes for the maximum mass 1.44~M$_\odot$. The solution of the stellar structure equations for a specific central density and central field strength provides a configuration with a mass and a radius. For a fixed field value of the field strength and a set of different central densities, we obtain the modified mass-radius relation. It shows that for a fixed central field as the central density reduces, the |$\mathcal{M}$/W| ratio increases and the M-R relation starts to deviate from the non-magnetic relation. As the effective magnetic energy increases the Lorentz force also increases which modifies the stellar structure to a non-spherical shape. Equilibrium structure with maximum density at the center can be obtained as long as the inward gravitational force dominates the Lorentz force component at any point within the star. The M-R relations with fixed |$\mathcal{M}$/W| value with either poloidal or toroidal field are similar to the non-magnetic M-R relation but shifted to a higher mass value. The maximum mass obtained for a pure poloidal field is about 1.9~M$_\odot$ \citep{Bera+Bhattacharya2014} whereas for a pure toroidal field the maximum mass lies beyond 5~M$_\odot$ \citep{Bera+Bhattacharya2016a} (Fig.~\ref{M_R}).

To study the effects of the modified EoS due to the quantized energy state of the electrons, we compute the equilibrium structure and compare with the configuration having the same central condition but with the degenerate EoS. We find that the modification due to quantized EoS is less than 1\% in their mass values. Similarly, the same procedure is followed to compare the effects of Salpeter EoS and general relativistic gravity on the equilibrium structures. The inclusion of these effects reduces the maximum mass just by a few percent which leads us to conclude that the equilibrium configurations obtained using degenerate EoS and Newtonian gravity are close to the exact solution \citep{Bera+Bhattacharya2016a}.

\articlefigure[width=.75\textwidth]{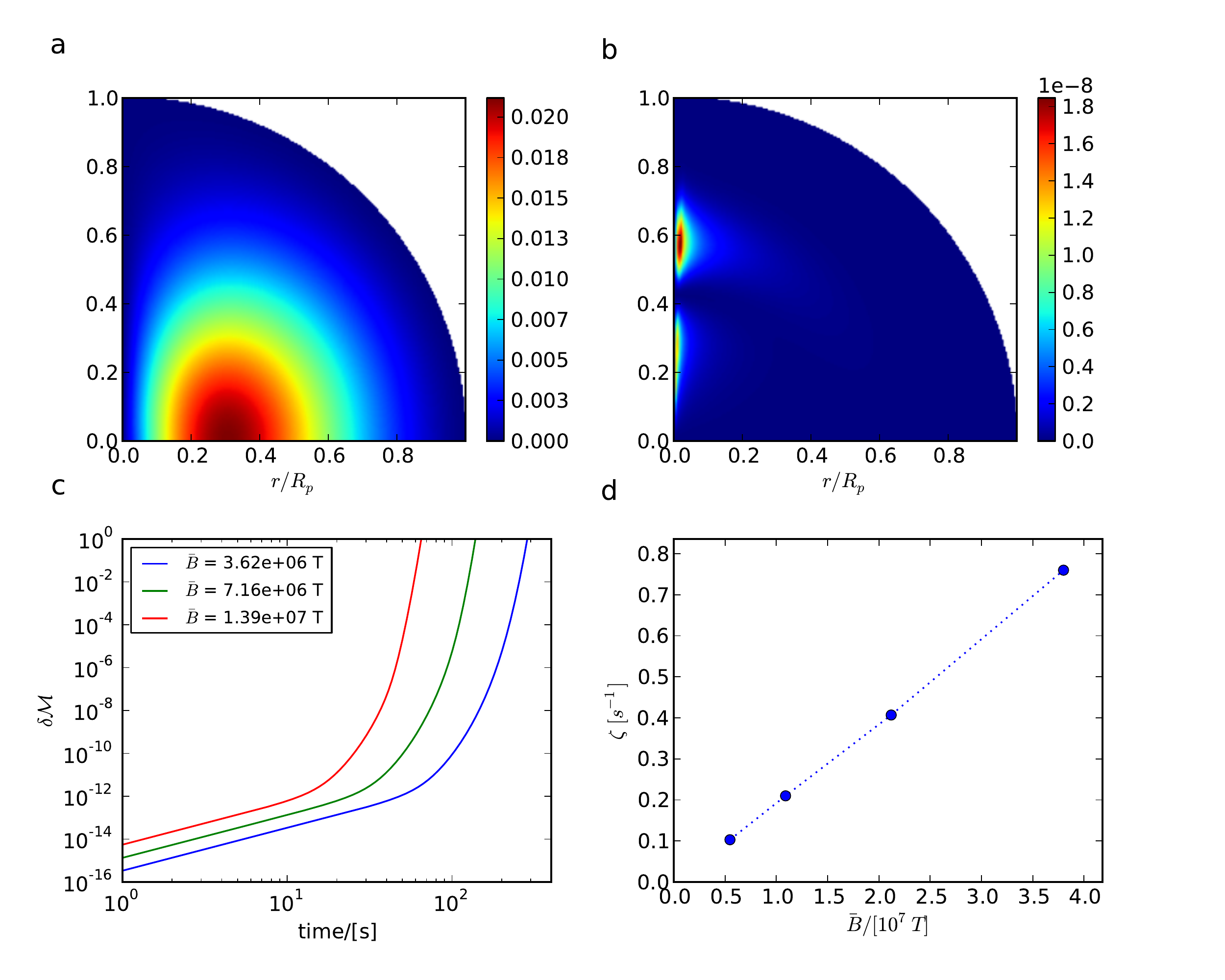}{lin_tor}{Linear perturbation study of white dwarfs with pure toroidal field: a) Toroidal field distribution of the equilibrium configuration with mass 0.88 M$_\odot$ and $\mathcal{M}/W$=0.8\%. b) The ratio of perturbed field magnitude to the equilibrium field after the onset of the instability for the $m = 1$ evolution. The instability appears near the axial region of this configuration. c) The evolution of perturbed field shows instability which appears corresponding to their Alfv\'en crossing time $\tau_A\sim 78, 39 ~\rm{and}~ 20~ s$. d) The growth rate ($\zeta$), at the beginning of the instability from the time evolution of Fig.~\ref{lin_tor}c, is almost linearly proportional to the average magnetic field strength.\label{lin_tor}}
The linear evolution of the pure toroidal field is studied with the velocity perturbation $\mathbf f \sim \hat{r} \times \nabla Y_{11}$ to the equilibrium solution and for the pure poloidal case we use $\mathbf f \sim \hat{r} \times \nabla Y_{22}$. The perturbation grows exponentially near the symmetry axis (or neutral line on the equatorial plane with vanishing field strength) for the pure poloidal (toroidal) case (Fig.~\ref{lin_tor}). The exponential growth of the perturbation, indicator of the instability, appears after about an Alfv\'en time ($\tau_A\approx R\sqrt{\frac{\mu_0\langle\rho\rangle}{\bar B^2}}$, where $R$ : stellar radius, $\langle\rho\rangle, \bar B$ are average density and field). The growth rate of the instability is inversely proportional to the average field strength. The non-linear evolution of the same kind of perturbation computed using \textsc{pluto} shows initially a similar  instability which later saturates (Fig.~\ref{nonlin_tor}). But the configuration with saturation deviates strongly from the initial, and the total field decays. This saturated state might not be exact as the fixed gradient boundary condition at the outer radial boundary and the Cowling approximation for the gravitational potential may not remain valid. A spherical harmonic decomposition of the perturbed field at a fixed radius shows lower order modes at the beginning but later, as instability grows, other higher order modes appear \citep{Bera+Bhattacharya2016b}. Fig.~\ref{M_Ralfven} shows that the massive configurations have a very short Alfv\'en time.

\articlefigure[width=.75\textwidth]{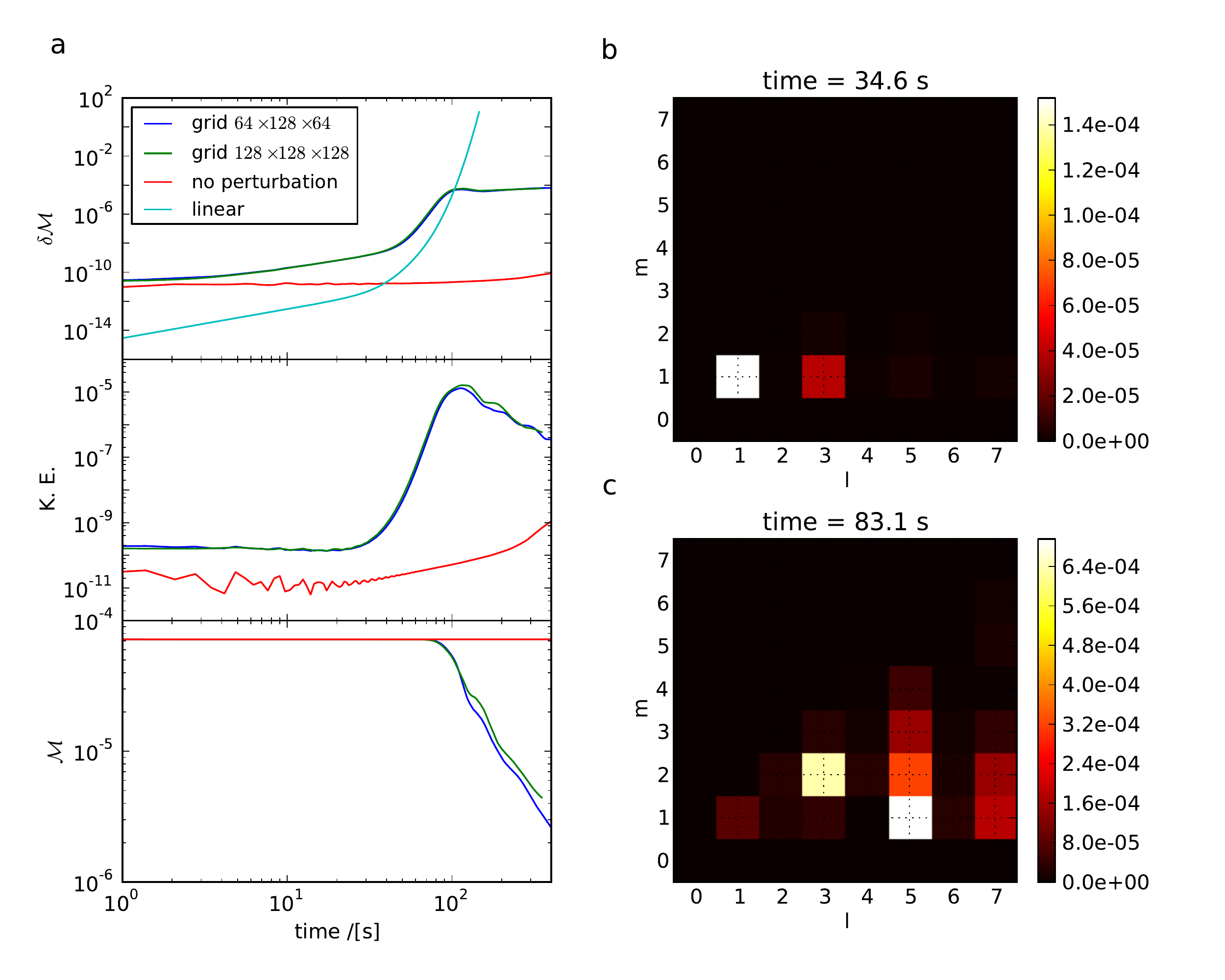}{nonlin_tor}{Non-linear perturbation study of white dwarfs with a pure toroidal field: a) The evolution of i) perturbed magnetic energy ($\delta \mathcal M$), ii) kinetic energy (K.E.) and iii) total magnetic energy ($\mathcal M$). The results are almost independent of the grid resolution as the value of the parameters for $64^3$ and $128^3$ are identical. The linear evolution of the $m=1$ perturbation mode of this polytropic star is shown for comparison. The non-linear evolution of the equilibrium configuration without any perturbation shows the static characteristics. b) Early time (t~=~35~s) spherical harmonic mode components of the $\theta$-component of the perturbed field ($\delta B_\theta$) evaluated at $r/R=0.6$. c) Late time (t~=~83~s) mode components of $\delta B_\theta$ at the same $r$ value.}
\articlefigure{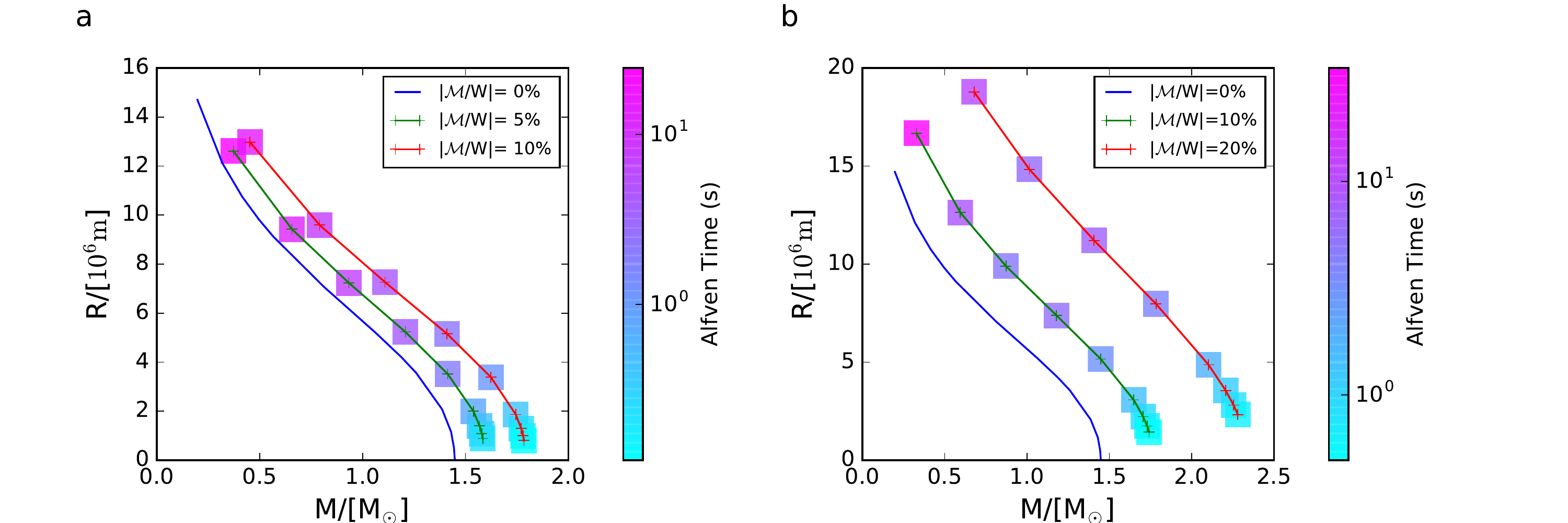}{M_Ralfven}{The mass-radius relation of white dwarfs with a) pure poloidal (left panel) and b) pure toroidal field. The coloured squares exhibit the Alfv\'en crossing time ($\tau_A$) of these configurations.}

\section{Conclusion}
We solve the magnetic axisymmetric stellar structure equations and present the mass-radius relation for the strongly magnetized configurations. We also validated the Newtonian framework with degenerate EoS. The linear and non-linear evolution studies of the perturbation configurations show the inherent short time-scale instability. The main conclusions are,
\begin{itemize}
 \item WD can support a larger mass in the presence of a strong magnetic field.
 \item Even at the maximum strength of the magnetic field, the impact of Landau quantization on the stellar structure is not significant.
 \item Long-lived super-Chandrasekhar WDs supported by the magnetic field are unlikely to occur in Nature.
\end{itemize}



\end{document}